\documentclass[letterpaper, 10 pt, conference]{ieeeconf}  %
\usepackage{graphicx}  
\usepackage[font=footnotesize, labelsep=period]{caption}
\usepackage{booktabs}        
\usepackage{amsmath} 
\usepackage{amssymb}
\usepackage{siunitx}                
\usepackage{array}             
\usepackage{tabularx}                
\usepackage[hidelinks]{hyperref}
\usepackage{threeparttable} 
\usepackage{float}
\usepackage{subcaption}
\IEEEoverridecommandlockouts                   
\title{\LARGE \bf
TeleSim: A Network-Aware Testbed and Benchmark Dataset for Telerobotic Applications}

\author{Zexin Deng, Zhenhui Yuan~\IEEEmembership{Senior Member,~IEEE}, and Longhao Zou%
\thanks{Z. Deng and Z. Yuan are with the School of Engineering, University of Warwick, Coventry CV4 7AL, United Kingdom (e-mail: \{zexin.deng, zhenhui.yuan\}@warwick.ac.uk).}%
\thanks{L. Zou is with the Pengcheng Laboratory, P.R. China (e-mail: zouhl@pcl.ac.cn).}}

\begin{document}
\maketitle
\thispagestyle{empty}
\pagestyle{empty}

\begin{abstract}
Telerobotic technologies are becoming increasingly essential in fields such as remote surgery, nuclear decommissioning, and space exploration. Reliable datasets and testbeds are essential for evaluating telerobotic system performance prior to real-world deployment. However, there is a notable lack of datasets that capture the impact of network delays, as well as testbeds that realistically model the communication link between the operator and the robot. This paper introduces TeleSim, a network-aware teleoperation dataset and testbed designed to assess the performance of telerobotic applications under diverse network conditions. TeleSim systematically collects performance data from fine manipulation tasks executed under three predefined network quality tiers: High, Medium, and Low. Each tier is characterized through controlled settings of bandwidth, latency, jitter, and packet loss. Using OMNeT++ for precise network simulation, we record a wide range of metrics, including completion time, success rates, video quality indicators (Peak Signal-to-Noise Ratio (PSNR) and Structural Similarity Index Measure (SSIM)), and quality of service (QoS) parameters. TeleSim comprises 300 experimental trials, providing a robust benchmark for evaluating teleoperation systems across heterogeneous network scenarios. In the worst network condition, completion time increases by 221.8\% and success rate drops by 64\%. Our findings reveal that network degradation leads to compounding negative impacts, notably reduced video quality and prolonged task execution, highlighting the need for adaptive, resilient teleoperation protocols. The full dataset and testbed software are publicly available on our GitHub repository: \url{https://github.com/ConnectedRoboticsLab} and YouTube channel: \url{https://youtu.be/Fz\_1iOYe104}.
\end{abstract}

\section{INTRODUCTION}

Teleoperation systems, which integrate communication, perception, and robotic control technologies, have emerged as critical solutions for addressing technical challenges across multiple domains. In healthcare, according to the Lancet Commission on Global Surgery, more than 5 billion people globally do not have timely access to safe and affordable surgical and anesthesia care~\cite{Meara2015}. This inequity is particularly acute in low- and middle-income countries, where specialized healthcare resources are highly centralized and rural populations frequently experience life-threatening delays. By enabling remote experts to manipulate physical systems through real-time video feedback and intuitive control interfaces, teleoperation technology has demonstrated broad applicability in telesurgery, disaster response, offshore robotics, and space exploration.

However, practical deployment of teleoperation systems is severely constrained by unstable and heterogeneous network environments. Fluctuations in latency, jitter, and packet loss significantly impair both visual perception and control responsiveness. This degradation leads to operational delays, reduced efficiency, and increased task failures. Despite evidence that network impairments prolong task completion times and elevate operator stress and error rates, most publicly available teleoperation datasets assume idealized network conditions, neglecting real-world constraints.

Existing research on teleoperation performance evaluation exhibits clear limitations. Most prior studies have focused on usability and task execution under ideal or stable network conditions. For instance, Rodríguez-Sedano et al.~\cite{Rodríguez-Sedano2024} examined how haptic feedback influences user behavior and task performance in collaborative teleoperation, providing insights into interface design and operator support. Similarly, Wang et al.~\cite{wang2025} proposed passivity-aware control strategies to preserve stability under time-varying delays. While these efforts are foundational, they often overlook the compounded effects of real-world network impairments—such as fluctuating bandwidth, variable latency, jitter, and packet loss—which critically impact both perception and control in practical telerobotic deployments.

Recent studies have begun to acknowledge the critical influence of network conditions on teleoperation. Schüler et al.~\cite{Schüler2022} emphasize the lack of standardized evaluation frameworks for assessing remote operation performance under varying network constraints, and propose a controlled testbed to address this need. In parallel, Tian et al.~\cite{tian2022} proposed a predictive motion segmentation and synthesis method based on GMM/HSMM models to mask network-induced latency in cloud-based teleoperation. While these contributions are methodologically innovative, they primarily focus on algorithmic strategies and lack publicly available datasets with annotated network impairments, limiting reproducibility and benchmarking across different systems.

Despite growing interest in teleoperation benchmarking, existing datasets remain limited in scope and realism.  
The DY-Teleop dataset~\cite{liang2017} provides latency statistics (mean 194\,ms) but omits critical network impairment factors such as packet loss, jitter, or perceptual video quality metrics.  
Rodríguez-Seda et al.~\cite{rodriguez2009} evaluated bilateral teleoperation performance under up to 50\% packet loss, but the associated dataset is not publicly available.  
OPEN TEACH~\cite{iyer2024}, though extensive in task diversity, lacks annotations on network conditions.  
Hashempour et al.~\cite{hashempour2020} contributed a domain-specific dataset focused on needle insertion, but without controlled or labeled variations in network conditions, its utility for networking research is limited.

To bridge this gap, we design and implement TeleSim for realistic teleoperation evaluation. This work makes the following key contributions:

\begin{enumerate}
    \item \textbf{Hardware-in-the-Loop Testbed:} We develop a testbed integrating a 6-DoF robotic arm, OMNeT++-based network simulation, and real-time video transmission to enable realistic teleoperation scenarios.
    
    \item \textbf{Comprehensive Dataset:} We conduct 300 fine-manipulation trials across three network quality tiers, capturing 15 synchronized metrics that span network configuration, observed performance, video quality, and task outcomes.
    
    \item \textbf{Quantitative Analysis:} Under severely degraded network conditions, video quality deteriorates substantially—PSNR decreases 48.0\% (from 24.23\,dB to 12.59\,dB) and SSIM decreases 20.3\% (from 0.930 to 0.741). Task performance exhibits steeper decline with completion time increasing 221.8\% (from 117.7\,s to 378.8\,s) and success rate dropping from 92\% to 28\%, revealing compounded effects of network degradation.
\end{enumerate}

The structure of the paper is as follows: Section II describes the hardware-in-the-loop testbed and experimental setup; Section III details the development of the dataset; Section IV presents the experimental results and analysis; and Section V concludes with key findings and future work.

\section{HARDWARE-IN-THE-LOOP TESTBED}
This section presents the experimental testbed for evaluating teleoperation performance under controlled network conditions. The platform integrates three core components: a physical hardware system, an OMNeT++-based virtual network topology, and a synchronized data acquisition workflow, enabling systematic assessment of network impairment effects on teleoperation.
\subsection{Physical Setup}
\begin{figure}[!t]
  \centering
  \includegraphics[width=0.48\textwidth]{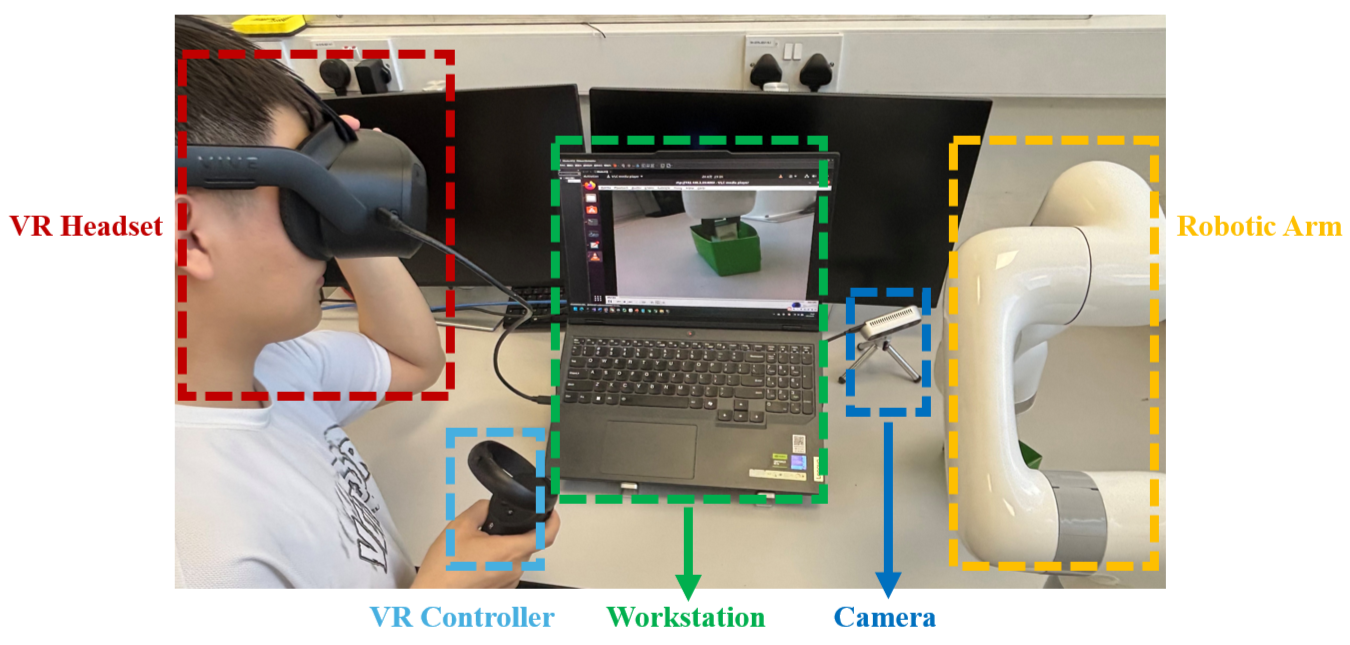}
  \caption{Illustration of the TeleSim testbed at the University of Warwick.}
  \label{fig:physical_setup}
\end{figure}

Figure~\ref{fig:physical_setup} shows the overall layout of the TeleSim platform. A RealSense D435i camera, mounted on a fixed tripod, captures RGB video of the robot workspace and streams it to a workstation running network simulation. The processed video is rendered in a VIVE Focus 3 VR headset, allowing the operator to perceive the task environment in real-time. Based on this visual feedback, the operator issues motion commands via a wireless VR controller, which are transmitted over a low-latency LAN to the robot controller. The current implementation uses a 6-DoF robotic arm controlled via the Robot Operating System (ROS). Due to its modular design, the platform supports rapid integration of other robotic hardware.

\subsection{Network Design}
\begin{figure}[!t]
\centering
\includegraphics[width=0.48\textwidth]{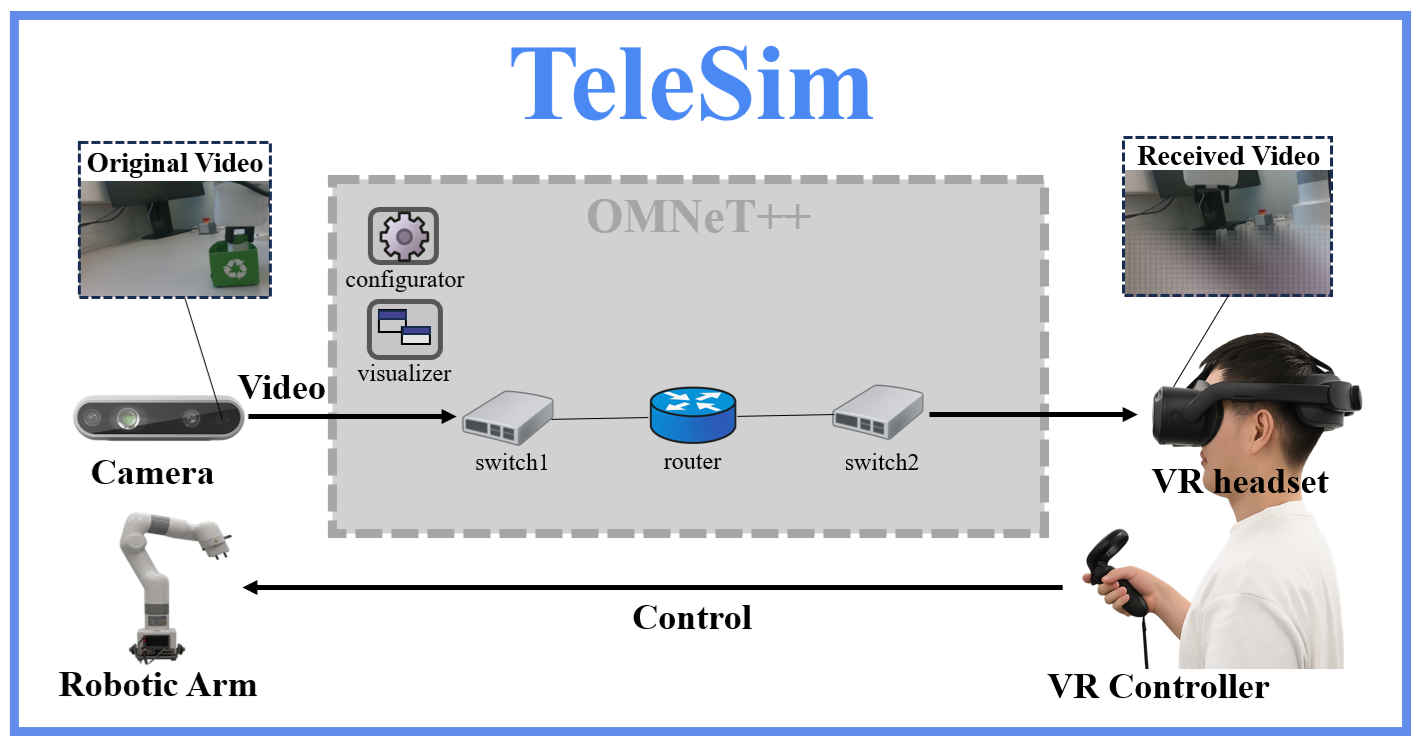}
  \caption{Hardware-in-the-loop network topology of the TeleSim testbed.}
  \label{fig:network_topology}
\end{figure}

\textbf{Network Topology:} Figure~\ref{fig:network_topology} illustrates the hardware-in-the-loop network topology of the TeleSim testbed. The RealSense camera captures video and streams it as RTP packets through a virtual network implemented in OMNeT++, consisting of two software-defined switches and a router. This network enables precise control over bandwidth, latency, jitter, and packet loss. The video is then delivered to the operator's VR headset. Control commands from the VR controller are transmitted via a dedicated Gigabit LAN directly to the robotic arm, ensuring low-latency and stable operation. Two example frames demonstrate video degradation under low-tier network conditions (10\,Mbps bandwidth, 1000\,ms latency, 200\,ms jitter, 3.0\% packet loss), highlighting the visual disparity between original and received streams.

\textbf{Network Conditions:}
Following ITU-T~Y.1541 classification guidelines and empirical Quality of Experience (QoE) thresholds~\cite{Dymarski2012,ITU-T}, we partition the 300 captured flow traces into three discrete network tiers: High, Medium, and Low. The parameters are summarized in Table~\ref{tab:nc_levels}.

\begin{table}[!t]
  \centering
  \caption{Various Network Conditions in line with ITU-T~Y.1541~\cite{ITU-T}}
  \label{tab:nc_levels}
  \small
  \renewcommand{\arraystretch}{1.15}
  \begin{tabular}{lccc}
    \toprule
    \textbf{Network Condition} & \textbf{High} & \textbf{Medium} & \textbf{Low} \\
    \midrule
    Bandwidth (Mbps)   & 1000  & 100   & 10   \\
    Latency (ms)       & 100   & 400   & 1000 \\
    Jitter (ms)        & 25    & 50    & 200  \\
    Packet Loss (\%)   & 0.1   & 1.0   & 3.0  \\
    \bottomrule
  \end{tabular}
\end{table}

\subsection{Experimental Design}
\textbf{Task Design:}  
A standardized pick-and-place task is defined to balance complexity and measurability. The task requires the operator to: (1) pick up a plastic block measuring $3.7 \times 2.1 \times 1.2$\,cm, located 20\,cm from the robot's initial pose; (2) place the block into a receptacle measuring $6 \times 5 \times 10$\,cm. A trial is considered successful only when both steps are completed without errors.

\textbf{Workflow:} 
Figure~\ref{fig:network_topology} outlines the data flow for each trial. 
The RealSense D435i captures RGB video at 640$\times$480 pixels and 30\,fps. The video stream is encoded using H.264 and transmitted as RTP/UDP into an OMNeT++ virtual network built with the INET framework, where predefined network impairments are applied. The operator views this stream in a VIVE Focus 3 headset and uses a VR controller to send motion commands via a clean Gigabit LAN to the Lite 6 robotic arm. A task-start trigger coincides with the camera's first frame, launching video recording and timer simultaneously. Upon task completion, the end trigger stops recording and stores elapsed time with success/failure flag. Recorded sequences are processed with MSU VQMT to compute frame-level PSNR and SSIM, with mean values serving as objective video quality indicators.

\section{BENCHMARK DATASET}\label{sec:dataset}

\subsection{TeleSim Overview}\label{subsec:overview}

TeleSim comprises 300 fine-manipulation trials performed under three predefined network quality tiers: High, Medium, and Low. These tiers emulate realistic communication constraints including limited bandwidth, variable latency, jitter, and packet loss. Each trial is conducted using a hardware-in-the-loop framework, capturing synchronized measurements of network behavior and task outcomes.

All trials follow a consistent task design and evaluation protocol to ensure comparability across conditions. The dataset captures comprehensive indicators covering network setup, observed performance, visual perception, and task execution. Details of the recorded parameters are provided in the following section.

\subsection{Dataset Structure and Parameters}
Table~\ref{tab:dataset_fields} details the 15 key parameters of TeleSim along with their units and descriptions, categorized into four main groups:

\textbf{Network Configuration Parameters:} Define the network conditions emulated in OMNeT++, including bandwidth limit, propagation delay, jitter, and packet loss rate. These parameters constitute the three network quality tiers (High, Medium, Low) defined in Table~\ref{tab:nc_levels}.

\textbf{Measured Network Metrics:} Reflect the actual network behavior observed during each trial, including mean and maximum values for throughput, one-way latency, jitter, and packet loss ratio. These metrics are collected in real-time using network monitoring tools (\texttt{nload}, \texttt{ifstat}, \texttt{tshark}), providing a dynamic view of network performance beyond preset configurations.

\textbf{Video Quality Metrics:} PSNR and SSIM quantify the visual fidelity of the operator's received video feedback. The VQMT tool computes frame-by-frame comparisons between sent and received videos, with mean values serving as indicators of overall video quality, directly reflecting the impact of network impairments on visual perception during teleoperation.

\textbf{Task Performance Metrics:} Include completion time and success flag, directly measuring the impact of network conditions on operational efficiency and success rate.

This structured approach enables researchers to explore the complex relationships between network conditions, video quality, and task performance, providing a foundation for network-aware design in telerobotic applications.

\begin{table}[!t]
  \caption{Structure and Metrics of the TeleSim Dataset}
  \label{tab:dataset_fields}
  \centering
  \footnotesize
  \setlength{\tabcolsep}{2pt}
  \renewcommand{\arraystretch}{1.05}
  \begin{tabularx}{\linewidth}{
    >{\centering\arraybackslash}p{0.45cm}
    >{\raggedright\arraybackslash}p{2.6cm}
    >{\centering\arraybackslash}p{0.75cm}
    X}
    \toprule
    \textbf{No.} & \textbf{Parameter} & \textbf{Unit} & \textbf{Description}\\
    \midrule
    \multicolumn{4}{c}{\textbf{Network Configuration in OMNeT++}}\\
    \midrule
    1  & Bandwidth & Mbps & Bandwidth limit\\
    2  & Latency   & ms   & Propagation latency\\
    3  & Jitter    & ms   & Variation of latency\\
    4  & Set PLR       & \%   & Packet loss ratio\\
    \midrule
    \multicolumn{4}{c}{\textbf{Measured Network Metrics}}\\
    \midrule
    5  & Mean Throughput & Mbps &  Mean received bitrate\\
    6  & Max Throughput  & Mbps & Maximum received bitrate \\
    7  & Mean Latency    & ms   & Mean one-way network delay\\
    8  & Max Latency     & ms   & Maximum one-way network delay\\
    9  & Mean Jitter     & ms   & Mean packet delay variation\\
   10  & Max Jitter      & ms   & Maximum packet delay variation\\
   11  & Measured PLR              & \%   & Packet loss ratio over capture period\\
    \midrule
    \multicolumn{4}{c}{\textbf{Video Quality Metrics}}\\
    \midrule
   12  & PSNR  & dB & Peak signal-to-noise ratio\\
   13  & SSIM  & –  & Structural similarity index measure\\
    \midrule
    \multicolumn{4}{c}{\textbf{Task Performance Metrics}}\\
    \midrule
   14  & Completion Time & s  & Time until task end\\
   15  & Success Flag  & –  & Binary task success\\
    \bottomrule
  \end{tabularx}
\end{table}


\section{RESULTS AND ANALYSIS}\label{sec:results}

\subsection{Impact of Network Quality on Video Quality}
\begin{figure}[!t]
  \centering
  \includegraphics[width=\linewidth]{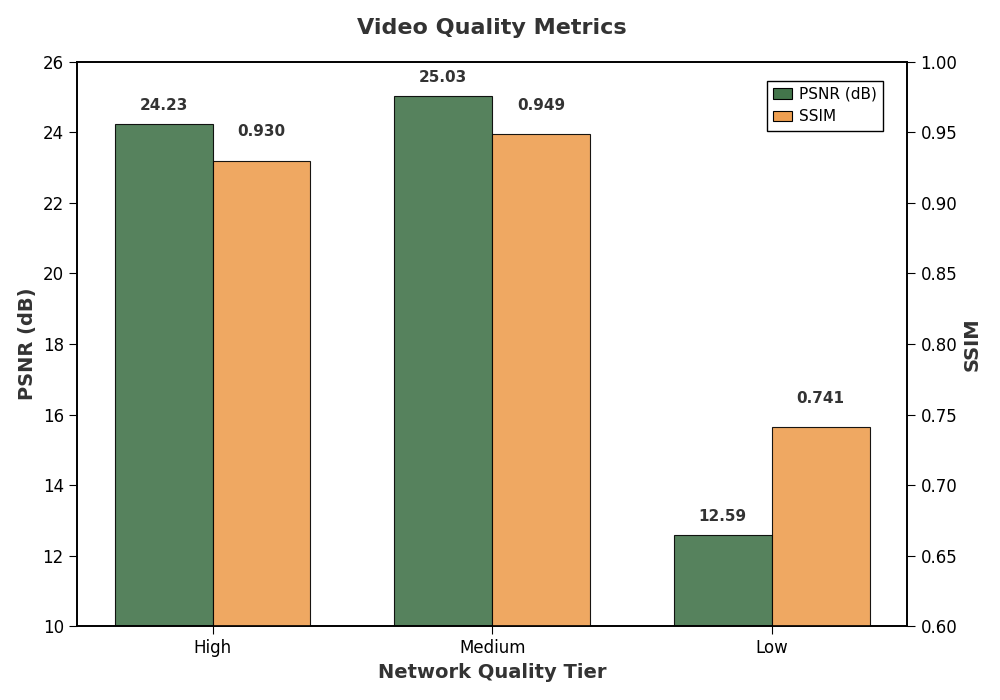}
  \caption{Average PSNR and SSIM for the three network tiers.}
  \label{fig:video_quality_metrics}
\end{figure}

Figure~\ref{fig:video_quality_metrics} presents the average PSNR and SSIM values across three network quality tiers. Results demonstrate clear degradation patterns as network conditions deteriorate. From High to Low tier, PSNR drops significantly from 24.23\,dB to 12.59\,dB (-48.0\%), while SSIM decreases from 0.930 to 0.741 (-20.3\%). Notably, the Medium tier achieves the highest video quality metrics (PSNR: 25.03\,dB, SSIM: 0.949), suggesting that adaptive streaming algorithms may optimize compression parameters more effectively under moderate network constraints.

\begin{figure}[!t] 
  \centering
  \subfloat[High network condition]{\includegraphics[width=0.30\linewidth]{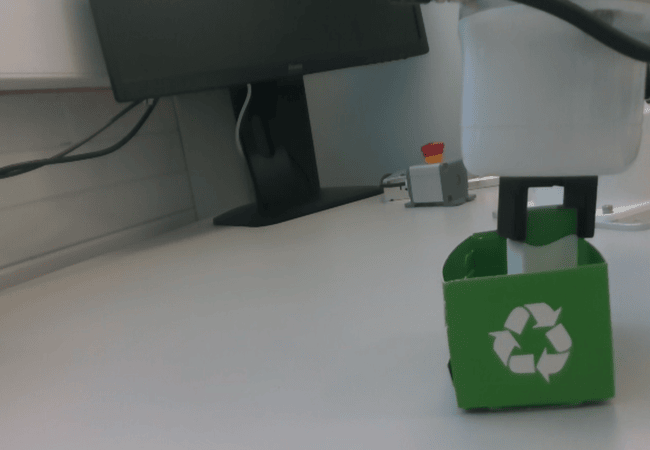}}\hfill
  \subfloat[Medium network condition]{\includegraphics[width=0.30\linewidth]{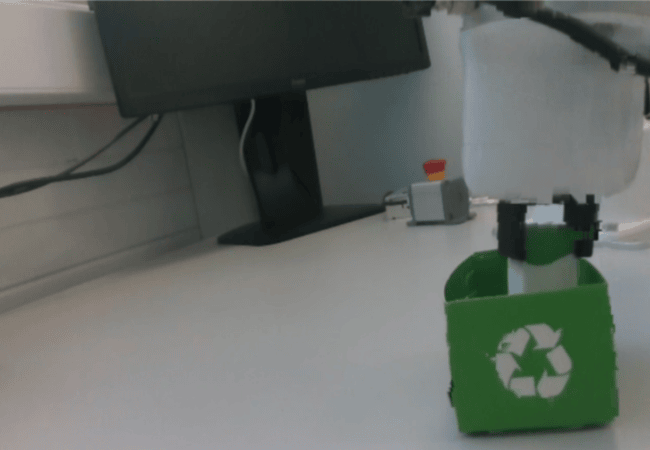}}\hfill
  \subfloat[Low network condition]{\includegraphics[width=0.30\linewidth]{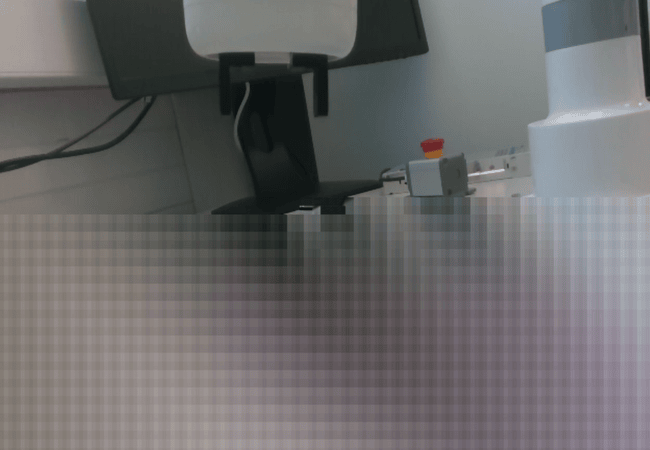}}
  \caption{Example video frames under the three network tiers.}
  \label{fig:frames}
\end{figure}

Figure~\ref{fig:frames} provides qualitative comparison from the operator's perspective. High tier (a) preserves sharp edges and vivid colors; Medium tier (b) exhibits mild blockiness and blur while maintaining object discernibility; Low tier (c) suffers from severe macro-blocking, motion smear, and intermittent freezes that significantly impair visibility of gripper tips and object edges.

\subsection{Impact of Network Quality on Task Performance}

\begin{figure}[!t]
  \centering
  \includegraphics[width=\linewidth]{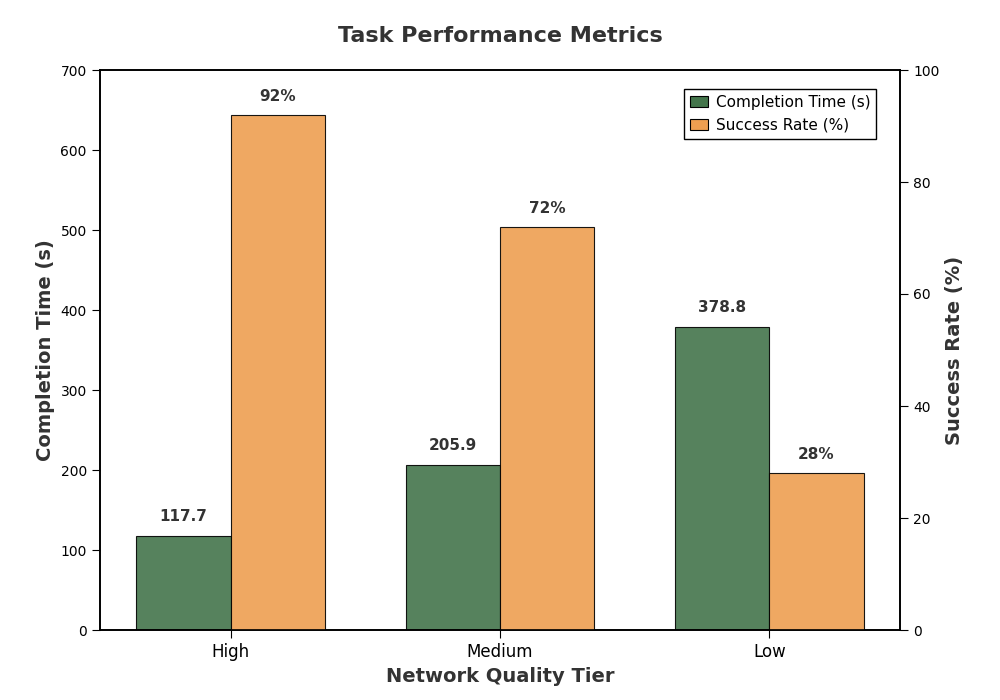}
  \caption{Average completion time and success rate for the three network tiers.}
  \label{fig:task_performance_metrics}
\end{figure}

Figure~\ref{fig:task_performance_metrics} illustrates the dramatic impact of network quality on operational efficiency. Average completion time increases from 117.7\,s in High tier to 378.8\,s in Low tier (+221.8\%). Success rate demonstrates steep decline from 92\% (High) to 28\% (Low), with Medium tier maintaining 72\% success at 205.9\,s average completion time.

These findings reveal a critical performance threshold: operators can adapt to moderate network degradation (75\% time increase) while severe degradation fundamentally compromises task feasibility. The non-linear relationship between network quality and task success underscores the importance of maintaining adequate network performance for telerobotic applications.

\section{CONCLUSIONS AND FUTURE WORK}

This paper introduces TeleSim, a network-aware hardware-in-the-loop testbed for teleoperation evaluation, along with a dataset of 300 fine manipulation trials across three network quality tiers. The system quantifies the impact of network impairments on both video quality and task performance.

Experimental results reveal dramatic performance degradation as network conditions deteriorate. Video quality metrics show PSNR dropping 48.0\% and SSIM decreasing 20.3\% from High to Low tier. Task performance exhibits even steeper decline with completion time increasing 221.8\% and success rate falling from 92\% to 28\%. These findings demonstrate the critical non-linear relationship between network quality and teleoperation performance.

\textbf{Limitations:} The current work focuses on RGB video-based teleoperation over wired connections. It does not account for multimodal data streams such as RGB-Depth (RGB-D) or haptic feedback, nor does it consider inter-operator variability or adaptive control schemes.

\textbf{Future Work:} Future development will advance in five directions: incorporating multimodal sensing modalities, simulating a wider range of communication links (e.g. cellular, satellite),  integrating real-world network trace datasets for increased realism, exploring more complex manipulation tasks and user interfaces, and enabling multi-user collaborative teleoperation scenarios.

\addtolength{\textheight}{-12cm}   





\end{document}